# Electron transport properties in high-purity Ge down to cryogenic temperatures

V. Aubry-Fortuna and P. Dollfus

*Institut d'Electronique Fondamentale, CNRS UMR 8622, Bât. 220, Université Paris-Sud, 91405 Orsay cedex, France.*

*Author contact: valerie.aubry-fortuna@u-psud.fr

**Abstract**

Electron transport in Ge at various temperatures down to 20 mK has been investigated using particle Monte Carlo simulation taking into account ionized impurity and inelastic phonon scattering. The simulations account for the essential features of electron transport at cryogenic temperature: Ohmic regime, anisotropy of the drift velocity relative to the direction of the electric field, as well as a negative differential mobility phenomenon along the <111> field orientation. Experimental data for the electron velocities are reproduced with a satisfactory accuracy. Examples of electron position in the real space during the simulations are given and evidence separated clouds of electrons propagating along different directions depending on the valley they belong.

# I.   Introduction

Germanium, as well as Si, exhibits a strong anisotropy of the electron effective mass tensor and a conduction band with several equivalent energy minima. As a consequence, the mean drift velocity in Ge and Si depends on field orientation relative to the crystal axes, especially below 190K.[1,2] Electronic transport in Ge has been extensively studied at temperatures down to 8K.[2] Today, with the development of cryogenic Ge detectors, knowledge of the carrier velocity laws at cryogenic temperatures becomes of utmost important.[3] However, few experimental data on electron transport in Ge at very low temperatures are reported in the literature[4-7] and the field range explored is rather limited (min. 0.6 V/cm, max. 4 V/cm in the <100> direction). Again, since the first results of Jacoboni *et al.*[2], theoretical studies have been mostly limited to the <100> field orientation[5,7], which prevents analysis of anisotropy effects.

In this context, this paper presents theoretical drift velocity versus electric field characteristics of electrons in Ge at temperatures down to 20 mK and along the <100> and <111> orientations. Section II presents the main features of our Monte Carlo code, in particular ionized impurity and inelastic acoustic phonon scatterings. Section III analyzes our results: strong anisotropy effects are demonstrated between the two field orientations studied here. A negative differential mobility phenomenon is found for the <111> orientation and explained in terms of the valley repopulation due to electron-phonon coupling. Concurrently, a comparison with existing experimental results is done. A rather good agreement is obtained between the experimental and the theoretical velocities in the field range explored. Finally, the consequence of the anisotropy of the transport at low temperature on the electron propagation in the real space is shown and discussed.

# II.   Monte Carlo model and simulation procedures

### *A.   Conduction-band structure of Ge*

The conduction band of Ge consists of four <111> L valleys at the edge of the Brillouin zone, one single <000> Γ valley at the zone centre and six <100> Δ valleys located near the zone edge. The lowest valleys are the L ones: the Γ valley and Δ valleys minima are 0.14 eV and 0.18 eV above, respectively. The equi-energy surfaces of L and Δ valleys are ellipsoidal with longitudinal ($m_l$) and transverse ($m_t$) effective masses. This surface is spherical for Γ valley. Details of effective mass values and other material parameters are reported in Table

I. In the present work, the non-parabolicity coefficient $\alpha$ is assumed to be equal to 0.3 eV$^{-1}$ for all valleys.[2]

### B. Scattering mecanisms

The scattering mechanisms included in the Monte Carlo model of electron transport in Ge are intra- and intervalley electron-phonon scattering by acoustic and optical phonons, and ionized impurity scattering.

The acoustic intravalley phonon scattering (ac) occurs in all valleys and is treated as an inelastic process, as described in Refs [2] and [8]. Indeed, at low temperature, the elastic approximation usually done for low energy acoustic phonons fails. The inelastic character of this scattering mechanism plays an important role in the electron relaxation.[9] In contrast to the case of elastic process, the exact expression of the number of phonons $N_q$ must be considered in the derivation of scattering rate, in order to treat correctly energy dissipation via acoustic phonon scatterings, i.e. $N_q = \left[\exp\left(\dfrac{\hbar q u}{k_B T}\right) - 1\right]^{-1}$, where $q$ is the phonon wave vector, $u$ is the longitudinal sound velocity, $k_B$ is the Boltzmann constant and $T$ is the lattice temperature. By defining the quantity $x$ proportional to the phonon wave vector $q$ as

$$x = \frac{\hbar q u}{k_B T}\sqrt{\frac{m_d}{m_0}} \tag{1}$$

the expression of the scattering probability $P_{ac}(\varepsilon,x)dx$ for ellipsoidal non-parabolic bands writes[8]

$$P_{ac}(\varepsilon,x)dx = \frac{\sqrt{m_d}(k_B T)^3 D_{ac}^2}{2^{5/2}\pi\hbar^4 u^4 \rho}\frac{1}{\sqrt{\gamma(\varepsilon)}}\left[\begin{array}{c}N_q(x)\\N_q(x)+1\end{array}\right](1+2\alpha\varepsilon \pm 2\alpha k_B T x)x^2 dx \tag{2}$$

with $\varepsilon$ the electron kinetic energy, $m_d = (m_l m_t^2)^{1/3}$ the density-of-states effective mass, $D_{ac}$ the acoustic deformation potential in eV, $\rho$ the material density and $\gamma(\varepsilon)$ equal to $\varepsilon(1+\alpha\varepsilon)$.

Upper (lower) terms and signs in Eq.(2) refer to phonon absorption (phonon emission). Details for final integration over $x$ to obtain the scattering rate $\lambda_{ac}(\varepsilon)$ can be found in Ref. [8]. Finally, the rejection technique is used to determine the final state $x$ after scattering, by considering the scattering probability $P_{ac}(\varepsilon,x)$ in an interval $[a, b]$ where $a$ and $b$ are the lower and the upper limits of the integral $\int_a^b P_{ac}(\varepsilon,x)dx$, respectively. From a pair of random numbers $(R_1, R_1')$ uniformly distributed in $[0,1]$, the random quantity $x_s = a + (b-a)R_1$ is selected as the $x$ final state if

$$\frac{P_{ac}(\varepsilon, x_s)}{P_{ac}(\varepsilon, b)} \geq R_1'. \tag{3}$$

Otherwise $x_s$ is rejected and a new pair of random numbers $(R_2, R_2')$ is generated. The procedure is repeated until Eq.(3) is satisfied. The final electron kinetic energy after scattering is $\varepsilon_f = \varepsilon_i \pm \hbar qu$ where $\varepsilon_i$ is the electron kinetic energy before scattering and $q$ is deduced from Eq.(1) with $x = x_s$. This scattering mechanism is isotropic and the scattering angle is randomly selected on a uniform distribution in the range $[0, 2\pi]$.

Derivation of scattering rate for optical intravalley (op) and intervalley (iv) phonons is very similar to the case of zero-order intervalley phonon scattering in silicon described elsewhere.[11] The following formulae correspond to the case of ellipsoidal non-parabolic bands.

Only electrons in <111> L valleys are subject to optical intravalley phonon scattering. Indeed, optical intravalley transitions are forbidden by selection rules in <000> Γ valley and along the <100> axis.[10] The scattering rate $\lambda_{op}(\varepsilon)$ is given by

$$\lambda_{op}(\varepsilon) = \frac{(m_d)^{3/2} D_0^2}{\sqrt{2}\pi\hbar^2 \rho\hbar\omega_{op}} \begin{bmatrix} N_q \\ N_q + 1 \end{bmatrix} \sqrt{\varepsilon \pm \hbar\omega_{op}} \times \sqrt{1 + \alpha(\varepsilon \pm \hbar\omega_{op})} \left[1 + 2\alpha(\varepsilon \pm \hbar\omega_{op})\right] \tag{4}$$

with $D_0$ and $\hbar\omega_{op}$ the deformation potential (in eV/cm) and phonon energy, respectively.

The considered intervalley processes are: two intervalley scatterings between L valleys, one intervalley scattering between L and Γ, one intervalley scattering between L and Δ, two intervalley scattering between Δ valleys and one intervalley scattering between Δ and Γ. The general expression of $\lambda_{iv}(\varepsilon)$ is the following :

$$\lambda_{iv}(\varepsilon) = \frac{Z_{iv}(m_d)^{3/2} D_0^2}{\sqrt{2}\pi\hbar^2 \rho\hbar\omega_{iv}} \begin{bmatrix} N_q \\ N_q + 1 \end{bmatrix} \sqrt{\varepsilon \pm \hbar\omega_{iv} + \Delta\varepsilon_{iv}} \sqrt{1 + \alpha(\varepsilon \pm \hbar\omega_{iv} + \Delta\varepsilon_{iv})} \\ \times \left[1 + 2\alpha(\varepsilon \pm \hbar\omega_{iv} + \Delta\varepsilon_{iv})\right] \tag{5}$$

where $Z_{iv}$ is the number of possible final valleys, $\Delta E_{iv}$ the intervalley energy transition and $D_0$ and $\hbar\omega_{iv}$ the deformation potential (in eV/cm) and phonon energy, respectively. As for optical intravalley processes, those scatterings are isotropic.

Scattering by ionized impurities (imp) is usually neglected in theoretical studies of electronic transport in ultra-pure Ge at cryogenic temperatures takes into account. Here, it is included in spite of a low impurity level. As will be shown below, this mechanism has a significant influence on the carrier velocity laws at low temperature under low field conditions. Ionized impurity scattering is treated via the screened Coulomb potential using a momentum-dependent screening length $L$ defined by Takimoto[12]: all details of the model may be found in Ref. [11]. Results are presented here for a net density of ionized impurities of $10^{10}$ cm$^{-3}$, which is the typical order of magnitude of the residual density of impurities in high-purity Ge.

All material and phonon scattering parameters used in the present study are reported in Table I. All scattering rates were calculated and stored in a look-up table prior to the simulation of electron transport, in an energy range between $10^{-10}$ eV and a few eV. For energies larger than 1 meV, this is done using a uniform discretization with a 1 meV energy step. For energies less than 1 meV, an energy discretization with geometrical progression (typically 50 points per decade) has been implemented in our Monte Carlo (MC) code to allow an accurate description of very low energy transfers. Fig. 1a shows the scattering rates in <111> L valleys as a function of energy $\varepsilon$ at 20 mK and illustrates the two domains of calculation. Note that the absorption rate of acoustic phonons is not plotted since it is extremely weak (Fig. 1b) at such a low temperature.

Monte Carlo simulations were performed for an infinite Ge crystal subjected to a uniform electric field $E$, which can be applied in any direction. Initially, a minimum of 1,000 electrons are located in the four lowest <111> L valleys (25% in each), with their coordinates in k-space randomly selected according to equilibrium conditions. The motion of each electron consists of many free flights between two randomly-selected scatterings. Every time-step of $10^{-2}$ ps, an instantaneous drift velocity is extracted. Other microscopic quantities of interest are extracted, either every time-step, and/or at the end of the simulation: in particular, the average electron kinetic energy, the average electron kinetic energy per valley, the final electron population of the different valleys, the total number of scatterings and the proportion of each different scattering, as well as the proportion of each different scattering per type of phonon.

## III. Results and discussion

The drift velocity was calculated over an extended range of field intensity (from 0.01 up to $10^4$ V/cm), with <100> and <111> field orientations and for temperature $T$ ranging between 20 mK and 240 K. The results are reported in Fig. 2 and Fig. 3 (continuous lines). The most striking features are (i) the anisotropy effect which increases when decreasing the temperature, and (ii) the negative differential mobility occurring when the field is applied along <111> axes at low temperature. These effects were previously observed experimentally and verified by means of Monte Carlo simulation in Si[1] and in Ge[2] for temperatures above 8 K. They are due to a repopulation among the valleys when the angle between the field direction and the valley axis is not the same for all valleys at the minimum of conduction band, e.g. when the field is parallel to <100> in Si or parallel to <111> in Ge. Since it has important consequences on the establishment of the steady-state regime and on the relevant value of drift velocity to be considered with regard to experimental conditions in cryogenic detectors, the main features of this repopulation effect deserves to be described prior to the detailed analysis of results and comparison with experimental data.

### A. Valley repopulation with field parallel to <111>

With an electric field parallel to the <111> axis, a valley repopulation phenomenon occurs at low temperature and low electric field, as a consequence of the difference of effective mass felt by electrons along the field direction according to the valley they belong to. This repopulation is directly related to intervalley transitions via phonons of energy $\hbar\omega_{iv1}$ or $\hbar\omega_{iv2}$ (27.6 and 10.3 meV, respectively). At very low temperature, the number of phonons is so small that absorption processes are extremely rare. Additionally, if the electric field is small enough too, the average electron energy may be much smaller than the intervalley phonon energies, which makes the emission process also very rare. The time necessary to observe the steady-state repopulation may thus become very long, to such a point that it may be larger than the typical transit time in a wafer.

The same phenomenon was shown to appear for electrons in Si with field oriented along the <100> direction, i.e. along the longitudinal axis of two of the six Δ valleys of the conduction band.[1] In Ge, with electric field oriented parallel to <111>, the conduction effective mass of electrons in the L valley, with longitudinal axis parallel to the field, is much higher than in other L valleys oriented along $<1\bar{1}1>$, $<\bar{1}11>$ or $<\bar{1}\bar{1}1>$. Electrons are thus heated by the field more rapidly in the latter valleys, called the "hot" valleys, than in the L <111> valley, the "cold" valley. Electrons of the "hot" valleys can thus gain the energy needed to experience intervalley transitions by phonon emission process more rapidly than electrons of the "cold" valley, which is the origin of the repopulation.

This phenomenon is illustrated in Fig. 4a, where the time evolution of the fraction of electron in the "cold" valley is shown for a wide range of electric field applied along <111> axis at $T = 20$ mK. At very low field (< 0.4V/cm), intravalley and intervalley relaxation mechanisms tend to be time-separated and the time necessary to reach the steady-state repopulation extends dramatically: it is even greater than 20 μs for a field of 0.1 V/cm or less. For $E \geq 10$ V/cm, the heating of electrons is fast enough for all intravalley and intervalley processes to take place at the same time. The steady-state regime is reached within a few nanoseconds or less, and the repopulation of the "cold" valley is incomplete and field-dependent. At intermediate fields (≈ 0.4 - 9 V/cm), the repopulation of the "cold" valley is field-independent and leads to a steady-state regime where almost all electrons reside in the "cold" valley. The consequence on the drift velocity is reported in Fig. 4b, where we plot the time evolution of both the fraction of electrons in the L <111> valley and the drift velocity averaged over all particles, for $E_{111} = 1$ V/cm. Starting at $t = 0$ with an equi-distribution of electrons among the four L valleys, the fraction of electrons in the "cold" valley remains very close to 25% for about 0.1 μs. This time interval corresponds to an intravalley quasi-equilibrium state: after a transient overshoot velocity lasting about 1 ns, intravalley phonon scattering generates a quasi-steady-state regime with an average velocity of $2.26 \times 10^6$ cm/s. In this regime, intervalley processes are scarce and do not influence the state of the system: very few electrons from "hot" valleys are able to transfer into the "cold" valley by phonon emission. However, beyond $t = 0.1$ μs, the total number of intervalley transfers is no longer negligible. The energy loss associated with phonon emission makes the return to "hot" valleys impossible. The system slowly evolves to the situation where all electrons reside in the "cold" valley which is the actual permanent steady-state. This transfer is accompanied by a velocity decrease to the final value of $5.3 \times 10^5$ cm/s.

This raises a crucial question regarding the relevant simulation time to be considered to extract the velocity, according to intended utilization of the results. For instance, the velocity-field characteristics obtained at different simulation times $t_{sim}$ are plotted in Fig. 5 for a field applied parallel to <111> at $T = 20$ mK and 190 K. At 20 mK, for an electric field range 0.4-10 V/cm, the extracted velocity appears strongly dependent on $t_{sim}$. To the end of using the velocity-field laws to analyze the charge signals collected at electrodes of cryogenic detectors, the simulation time should be chosen equal to the charge collection time. Within the standard operation conditions of such detectors, the charge collection time is typically close to 1 µs.[6] Throughout this paper, this value of 1 µs will be considered as the reference time $t_{sim}$ for extracting the $v_e$($E_{111}$) curves as those plotted in Fig. 3.

### B.  Temperature dependence

Above 8 K, the drift velocity is temperature-dependent, whatever the field intensity and orientation. Below 8 K, a saturation phenomenon takes place for fields around and higher than 10 V/cm. Below 1 K, the velocity-field laws become essentially temperature-independent: results obtained at 1 K and 20 mK are superimposed. Fig. 1b represents the variations of scattering rates versus temperature for an electron energy of $10^{-2}$ eV. The temperature-independence of the velocity appears when the scattering processes involving acoustic phonon absorption become negligible.

### C.  Ohmic regime and mobility

At sufficiently low field depending on the temperature, the drift velocity is isotropic, i.e. independent of the field orientation, and the Ohmic regime is reached (as an example, see Fig. 5 for $E \leq 0.4$ V/cm at 20 mK or $E \leq 400$ V/cm at 190 K ). Below 4 K, ionized impurity scatterings are not negligible (Fig. 1b. Indeed, not taking this scattering process into account leads to overestimated velocity up to $\approx$ 1 V/cm. In such conditions the Ohmic regime can not be reached, as shown in Fig. 2 and Fig. 3 for temperatures of 20 mK and 1 K (curve indicated by the arrow). Above 4 K, acoustic phonon scattering always dominates over ionized impurity scattering and the Ohmic regime is systematically reached, whether impurity scattering is taken into account or not.

In this regime, the drift velocity $v_e$ is strictly proportional to the field. The resulting mobility is plotted in Fig. 6 as a function of temperature. The solid line shows the present theoretical results, while symbols correspond to some available experimental data above 4 K.[2, 13-17] The agreement between these different results is good.

### D.  Anisotropy of the drift velocity

As the field increases from the Ohmic regime, the anisotropy of the drift velocity is observed. As previously mentioned, the anisotropy becomes increasingly large as the temperature decreases, due to the valley repopulation phenomenon.

Fig. 7a shows the electron population in the different L valleys at 20 mK as a function of the electric field: closed and open symbols are for $E_{100}$ and $E_{111}$, respectively. The simulation time $t_{sim}$ considered here is 1 μs, as discussed in sub-section III.A. For $E_{100}$, the electrons have the same average kinetic energy whatever the valley they belong to (see Fig. 8a, closed symbols) and the various intervalley transitions are well balanced (see Fig. 9a, closed symbols). Electrons are then uniformly distributed in the four L valleys, whatever the electric field value. However, this population becomes lower than 25% at high $E_{100}$ (> 1000 V/cm) when electrons can jump into the upper X and Γ valleys. For $E_{111}$, as soon as the repopulation of the "cold" valley occurs, it favours a transport of electrons with a larger effective mass, and then a smaller velocity, which leads to the separation of the $v_e(E_{100})$ and $v_e(E_{111})$ curves (see Fig. 5). The electron over-population in the "cold" valley *versus* $E_{111}$ goes through a maximum, from which intervalley transitions from the "cold" valley are no longer negligible (see Fig. 9a, open symbols) since electron kinetic energy progressively increases with the field (see Fig. 8a, open symbols).

Fig. 7b shows the electron population in the "cold" valley versus $E_{111}$ for various temperatures between 20 mK and 190 K. When increasing the temperature, the electron over-population effect becomes less and less important and the maximum is shifted towards higher electric field. At 190 K, the electron kinetic energy, in any valley and whatever the field orientation, is equal to the thermal energy up to about 100 V/cm (see Fig. 8b) and intervalley transitions are well balanced (see Fig. 9b). "Cold" and "hot" valleys can be distinguished only for high electric fields $E_{111}$, when electron kinetic energies are larger than the thermal energy $\frac{3}{2}k_BT$. Nevertheless, the difference in the number of intervalley transitions is rather weak and the over-population in the "cold" valley is limited. Fig. 5 shows that the anisotropy effect of the $v_e$ curves at 190K is then much less pronounced than at 20 mK.

### E.   *Negative differential mobility*

A related phenomenon of the anisotropy of the drift velocity is the occurrence of a negative differential mobility regime in the <111> field orientation. Fig. 3 clearly evidences this phenomenon at 20 K and below. Within a limited field range (typically between 1 V/cm and 10 V/cm), the transfer of electrons into the "cold" valley is of such magnitude (more than 80% according to Fig. 7b) that the averaged drift velocity over the different valleys actually decreases as the field increases, hence the negative mobility effect. The maximum of population in the cold valley then corresponds to a local minimum in the drift velocity. As the field further increases, due to a gradual equalization of the electron population in the different valleys as explained in sub-section III.D, the averaged drift velocity increases again.

## F. Comparison with experimental and other theoretical results

Our simulations reproduce with a rather good accuracy the temperature and field dependencies of the drift velocity, as obtained from earlier available experimental and theoretical studies above 8K.[2] Fig. 2 and Fig. 3 show these literature results for comparison with our simulations: the experimental results and the theoretical ones are marked by symbols and by dotted lines, respectively. Unfortunately, for a <111>-oriented field, experimental results at 8 K are missing below 80 V/cm and the negative differential mobility area can not be evidenced. Nevertheless, this phenomenon has been observed under specific experimental conditions: when enhancing the electron concentration up to $3\times10^{13}$ cm$^{-3}$ by tuning the intensity of the electron beam, time-of-flight measurements were performed down to 1 V/cm.[2] A negative differential mobility area then appears, but the corresponding results can not be directly compared to our simulations. Indeed, for such a high concentration, electron-electron scattering (not included in our simulations) is likely to influence the results significantly by tending to equalize the energy of electrons between "hot" and "cold" valleys.[2]

Fig. 10 presents the experimental results of Refs [4,6,7] and the corresponding theoretical curve $v_e(E_{100})$ from our Monte Carlo simulations at 20 mK (with ionized impurity scattering). The set of experimental data is obtained in a limited range of field intensity (0.06 to 4 V/cm). Results of an earlier theoretical study by Monte Carlo simulation[5] are also plotted. The latter is based on different and simplifying assumptions: the conduction band anisotropy was averaged out and ionized impurity scattering was not taken into account. The experiment/theory agreement appears to be better with our values of drift velocity, since they are lying in-between the experimental values of Refs [6] and [7]. This agreement is excellent regarding the variation rate of the velocity as a function of the field. The weak discrepancy between the experimental results may arise from the approximations made in extracting the carrier velocities, or from the "high-purity" character of Ge, i.e. the residual level of ionized impurities in the Ge substrate used for time-of-flight measurements. From the theoretical point of view, the screening effect is known to be difficult to consider accurately in the treatment of electron-impurity scattering and the simplified screening model currently used (see section II.B) may also be suspected to be at the origin of a discrepancy. In the case of a <111>-oriented field, experimental data are unfortunately not available for comparison with our theoretical results.

Finally, the carrier velocity-field laws obtained at 20 mK were used to model the charge collection process and to analyze the ionization signals in coplanar grid cryogenic Ge detectors for the Edelweiss dark matter research experiment.[18] Experimental and theoretical time-resolved ionization signals were successfully superimposed (Fig. 3 of Ref. [18]), which suggests that our model correctly describes the electron transport in Ge, including all specific features related to cryogenic temperatures.

### G. Electron propagation in the real space

A noteworthy phenomenon, directly related to anisotropic effects in Ge, concerns the trajectories of electrons in the real space, especially at very low temperature and low field. A first example is reported in Fig. 11a, in the case of $E_{100}$ = 1 V/cm at 20mK and $t_{sim}$ = 100 ns. This figure shows four clouds of electrons (one colour for one valley) in a plane perpendicular to the propagation direction [100]. Under these conditions of field and temperature, the electrons in the four valleys separate in the real space and propagate along different directions depending on the valley they belong to. This phenomenon occurs also below 1V/cm and is due to the low level of intervalley scattering. According to Fig. 9a, for $E_{100}$ lower than or equal to 1V/cm, the number of intervalley scattering is balanced between the different valleys, but it is very small, typically less than 18 during 100 ns for 1,000 electrons. With such low level of intervalley transitions, each electron remains confined into a single valley during its motion. As soon as the number of intervalley scattering increases, i.e. for $E_{100}$ = 10 V/cm ($10^5$ scatterings), all electrons propagate together within a single packet, as it can be seen in the inset of Fig. 11a. These results are in accordance with those of Ref. [7] regarding the oblique propagation of electrons in Ge.

Another example of electron propagation is given in Fig. 11b for $E_{111}$ = 1 V/cm at 20 mK and $t_{sim}$ = 100 ns. In the plane perpendicular to the propagation direction [111], four separated clouds of electrons follow again four different quasi-ballistic trajectories. However, only three of them present an oblique propagation, as defined previously, and the fourth cloud (corresponding to the "cold" valley) propagates along the electric field, i.e. along its longitudinal axe (blue symbols in Fig. 11b). As expected and as shown in the inset of Fig. 11b, the electrons of the "cold" valley move more slowly than all the others. For greater simulation times, due to the repopulation phenomenon in the case of <111>-oriented electric field (see Fig. 4a), only this latter cloud of electrons exists and then continues to propagate.

Obviously, these latter results are of great importance for a better understanding of charge collection processes in cryogenic Ge detectors, especially regarding the asymmetries observed in hole and electron collection.[19]

# IV. Conclusion

In this paper, a detailed theoretical investigation of electron transport in Ge at various temperatures has been presented. By means of Monte Carlo simulation, the transport properties have been analyzed for an electric field applied along the <100> and <111> directions by taking into account in particular ionized impurity and inelastic phonon scatterings. The well-known anisotropy of the drift velocity relative to the direction of the electric field was clearly evidenced, as well as a negative differential mobility effect when applying an electric field along <111>. These particularities were interpreted in terms of valley repopulation phenomenon, which was of great importance for determining the relevant velocity-field laws at low temperature and was then pointed out in the present paper. The Ohmic region was always reached whatever the temperature and mobility values were extracted and successfully compared to experimental ones. More generally, the simulations reproduced with a rather good accuracy the available experimental data. Finally, some examples of electron propagation in the real space evidenced the valley-dependence of quasi-ballistic trajectories under field and temperature conditions of negligible intervalley scattering.

# V. Acknowledgments

The authors are grateful to A. Broniatowski for fruitful discussions and continuous encouragement during the course of this work.

# VI. References

[1] C. Canali, C. Jacoboni, F. Nava, G. Ottaviani, A. Alberigi-Quaranta, Phys. Rev. B **12**, 2265 (1975).

[2] C. Jacoboni, F. Nava, C. Canali, G. Ottaviani, Phys. Rev.B **24**, 1014 (1981).

[3] A. Broniatowski, Nucl. Instrum. Methods A **520**, 178 (2004).

[4] B. Censier, "Étude et optimisation de la voie ionisation dans l'expérience Edelweiss de détection directe de la matière noire" (text in French), PhD thesis, University of Paris-Sud, Orsay, France, 2006.

[5] K.M. Sundqvist, B. Sadoulet, J. Low Temp. Phys. **151**, 443 (2008).

[6] V. Aubry-Fortuna, A. Broniatowski, P. Dollfus, Proceedings of the 13th International Workshop on Low temperature Detectors, AIP Conference Proceedings Vol. **1185**, 635 (2009).

[7]B. Cabrera, M. Pyle, R. Moffatt, K.M. Sundqvist and B. Sadoulet, arXiv:1004.1233v1 (2010) [astro-ph.IM].

[8]C. Jacoboni, P. Luigi, in The Monte Carlo method for semiconductor device simulation, edited by S. Selberherr (Springer-Verlag, Wien, New York), 1989.

[9]F. Hesto, J.-L. Pelouard, R. Castagné, J.-F. Pône, Appl. Phys. Lett. **45**, 641 (1984).

[10]W.A. Harrison, Phys. Rev. **104**, 1281 (1956).

[11]V. Aubry-Fortuna, P. Dollfus, S. Galdin-Retailleau, Solid-State Electronics **49**, 1320 (2005).

[12]N. Takimoto, J. Phys. So. Jpn. **14**, 1142 (1959).

[13]F.J. Morin, Phys. Rev. **93**, 62 (1954).

[14]E.M. Conwell, P.P. Debye, Phys. Rev. **93**, 693 (1954).

[15]M.B. Prince, Phys. Rev. **130**, 2201 (1963).

[16]R.D. Brown, S.H. Koening, W.Schillinger, Phys. Rev. **128**, 1688 (1962).

[17]R.S. de Biasi, S.S. Yee, J. Appl. Phys. **43**, 609 (1972).

[18]A. Broniatowski, X. Defay, E. Armengaud, L. Bergé, A. Benoit *et al.*, Physics Letter B **681**, 305 (2009).

[19]A. Broniatowski, private communication.

**FIGURE CAPTIONS**

Fig. 1: (a) Example of scattering rates as a function of electron energy ε for electron-phonon and electron-ionized impurity scatterings at 20 mK in L <111> valleys : "imp" stands for for ionized impurity, "ac" for acoustic intravalley, "op" for optical intravalley and "iv1, iv2" for the two L ↔ L intervalley processes. Values represented in dotted lines have been calculated using an energy discretization in a geometrical progression. (b) Variations of some scattering rates with temperature for an electron energy of $10^{-2}$ eV. In (a) and (b), letters "_e" and "_a" refer to for emission and absorption, respectively.

Fig. 2: Monte Carlo simulations of the field-velocity curves $v_e(E_{100})$ for temperatures ranging from 20mK to 240K (in continuous lines). Symbols and dotted lines are for experimental and theoretical results from Ref. [2], respectively.

Fig. 3: Monte Carlo simulations of the field-velocity curves $v_e(E_{111})$ for temperatures ranging from 20 mK to 240 K (in continuous lines). Symbols and dotted lines are for experimental and theoretical results from Ref. [2], respectively. Note the difference in the field scales below 8K (bottom) and above 20K (top).

Fig. 4: (a) Electron population (in %) in the L <111> valley as a function of the simulation time (in μs), when applying different field strengths in the <111> direction (0.1, 1, 4, 10 and 100 V/cm) at 20 mK. (b) Time variation of the L <111> population (in red) and of the averaged drift velocity (in blue) for $E_{111}$ = 1 V/cm and $T$ = 20 mK. The curves evidence four different regimes: the transient regime, the quasi steady-state regime, the repopulation regime and the steady-state regime.

Fig. 5: Velocity laws at 20 mK and 190 K for various simulation times $t_{sim}$ (0.02, 1 and 10 μs) when applying an electric field in the <111> direction.

Fig. 6: Ohmic electron mobility in Ge as a function of temperature. The solid line indicates the Monte Carlo mobility. Symbols refer to experimental data: closed triangles for Jacoboni[2], open circles for other data.[13-17]

Fig. 7: (a) Electron population (in %) at 20 mK in the different L valleys as a function of the electric field. Closed symbols are for $E_{100}$ and open symbols for $E_{111}$. In the latter case, results for L <1$\bar{1}$1> and L <$\bar{1}$ $\bar{1}$1> valleys are not reported, since they are identical to those plotted for L <$\bar{1}$11>. (b) Variation of the electron population (in %) in the L <111> valley as a function of $E_{111}$ for various temperatures. For all these results, $t_{sim}$ = 1 μs.

<017> Electron transport properties in high-purity Ge down to cryogenic temperatures

Fig. 8: Variation of the electron kinetic energy in the four L valleys as a function of the electric field at: (a) 20 mK and (b) 190 K. Closed symbols are for $E_{100}$ and open symbols for $E_{111}$. In the latter case, results for L $<1\bar{1}1>$ and L $<\bar{1}\bar{1}1>$ valleys are not reported, since they are identical to those plotted for L $<\bar{1}11>$.

Fig. 9: Example of variation of the number of intervalley scatterings per second and per electron as a function of the electric field: (a) at 20 mK and (b) at 190 K. Closed symbols are for $E_{100}$ and open symbols for $E_{111}$.

Fig. 10: Comparison between experimental[4, 6, 7] and theoretical[5, this work] velocity laws at 20mK for a <100>-oriented electric field.

Fig. 11: Examples of position in the real space of 1000 electrons under a <100> (a) or a <111> (b) applied electric field at 20 mK during $t_{sim}$= 100 ns.

**TABLES**

Table I. Material and phonon scattering parameters for electrons in Ge, according to Ref. [2].

| Lattice constant $a_0$ (Å) | 5.657 | Valleys L | $m_l$ = 1.588 $m_0$ |
|---|---|---|---|
| Density $\rho$ (kg/m$^3$) | 5.327 × 10$^3$ | | $m_t$ = 0.082 $m_0$ |
| Longitudinal sound velocity $u$ (m/s) | 5.4 × 10$^3$ | Valley $\Gamma$ | $m_{eff}$ = 0.037 $m_0$ |
| Dielectric constant | 16.2 | Valleys $\Delta$ | $m_l$ = 1.353 $m_0$ |
| Band gap (eV) | 0.66 | | $m_t$ = 0.288 $m_0$ |

| Acoustic intravalley | Deformation Potential $D_{ac}$ (eV) |
|---|---|
| in <111> L valleys | 11 eV |
| in <000> $\Gamma$ valleys | 5 eV |
| in <100> $\Delta$ valleys | 9 eV |

| Optical intravalley | Deformation Potential $D_{op}$ (eV/cm) | Phonon energy $\hbar\omega_{op}$ (meV) |
|---|---|---|
| in <111> L valleys | 5.5 × 10$^8$ | 37 |

| Intervalley | Deformation Potential $D_{iv0}$ (eV/cm) | Phonon energy $\hbar\omega_{iv}$ (meV) |
|---|---|---|
| L ↔ L (iv1) | 3.0 × 10$^8$ | 27.6 |
| L ↔ L (iv2) | 2.0 × 10$^7$ | 10.3 |
| L ↔ $\Gamma$ (iv3) | 2.0 × 10$^8$ | 27.6 |
| L ↔ $\Delta$ (iv4) | 4.06 × 10$^8$ | 27.6 |
| $\Delta$ ↔ $\Delta$ (iv5) | 9.46 × 10$^8$ | 37 |
| $\Delta$ ↔ $\Delta$ (iv6) | 7.89 × 10$^7$ | 8.6 |
| $\Delta$ ↔ $\Gamma$ (iv7) | 1.0 × 10$^9$ | 27.6 |

Figure 1

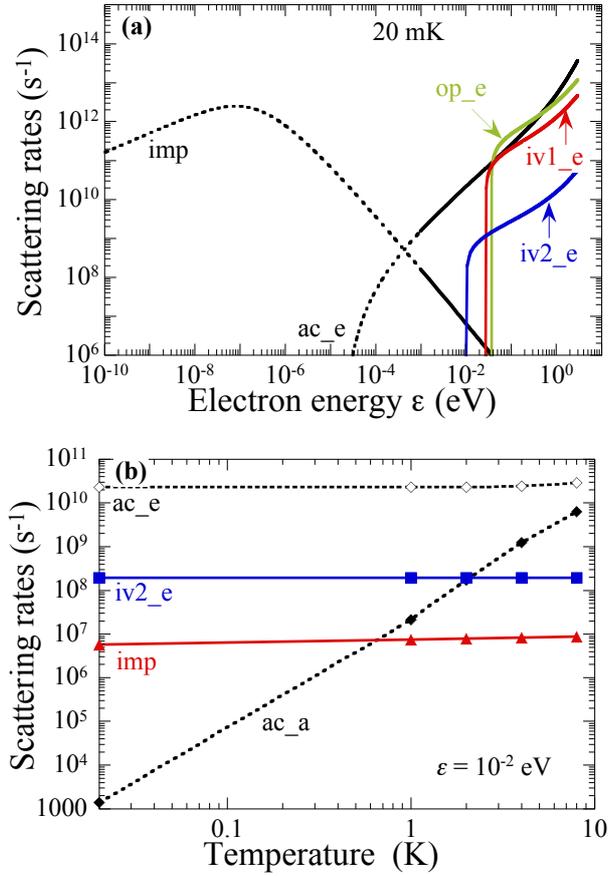

Figure 2

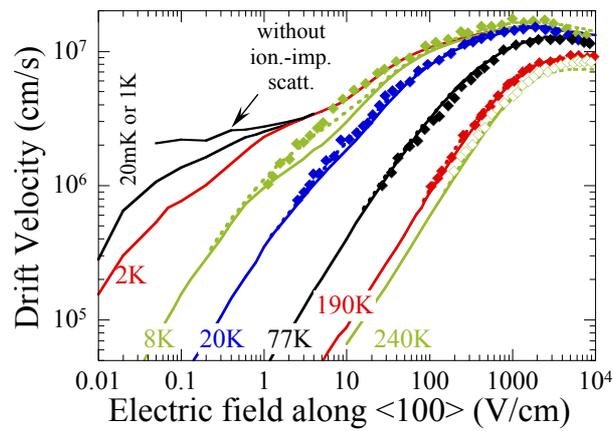

Figure 3

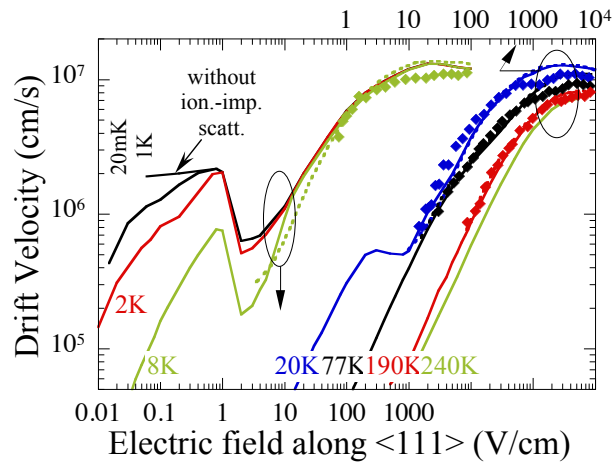

Figure 4

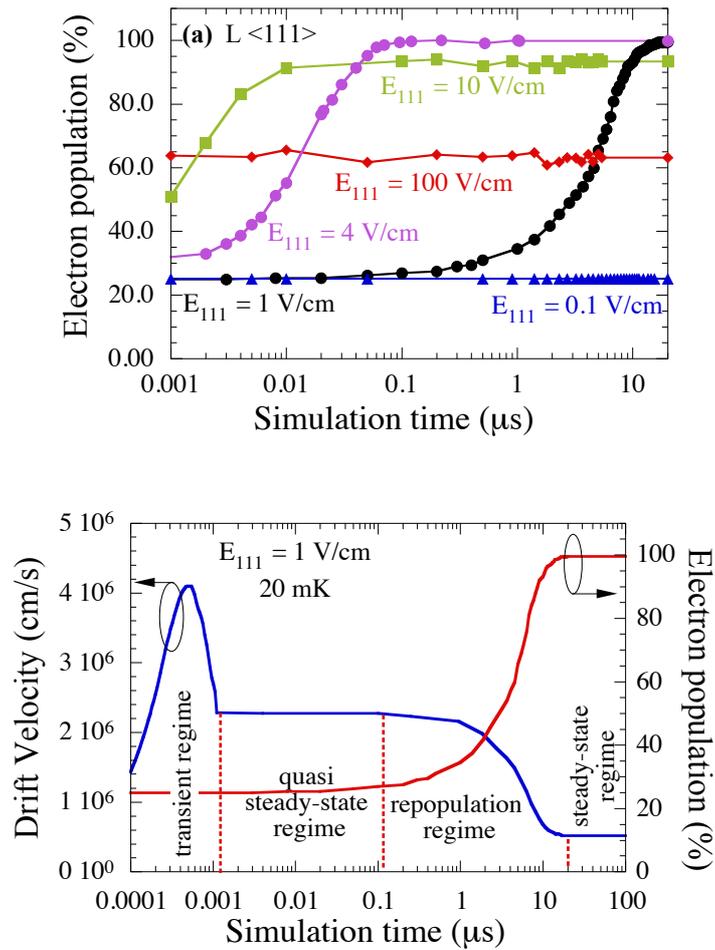

Figure 5

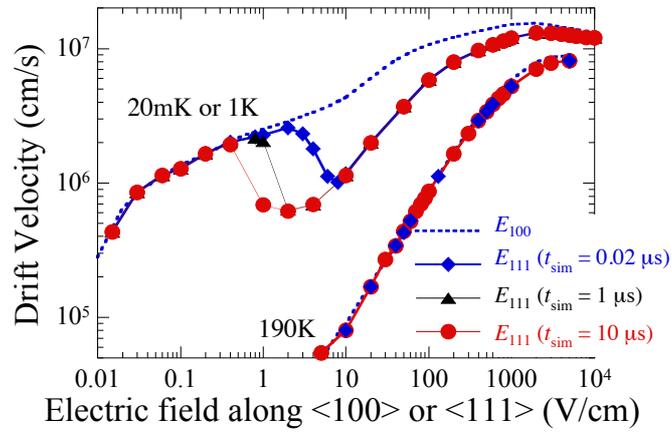

Figure 6

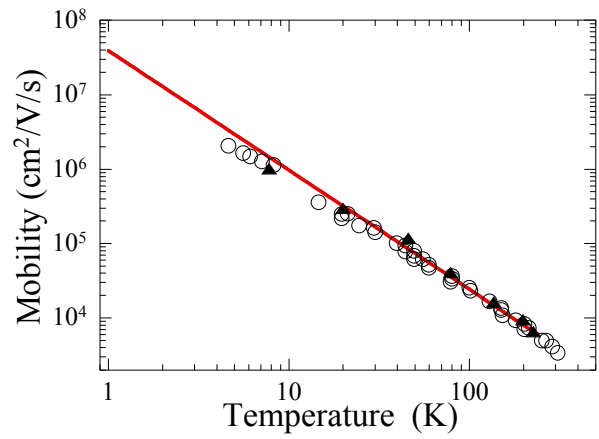

Figure 7

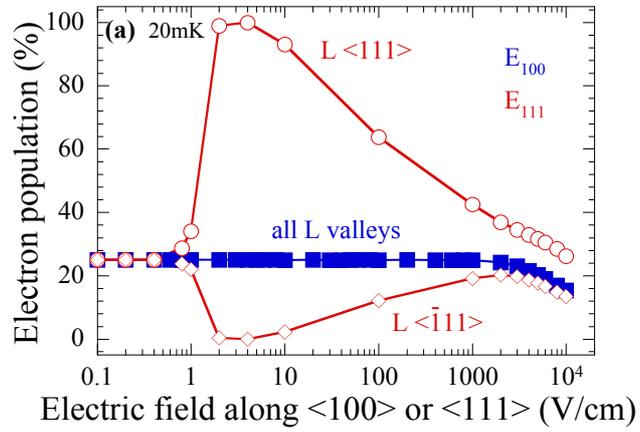

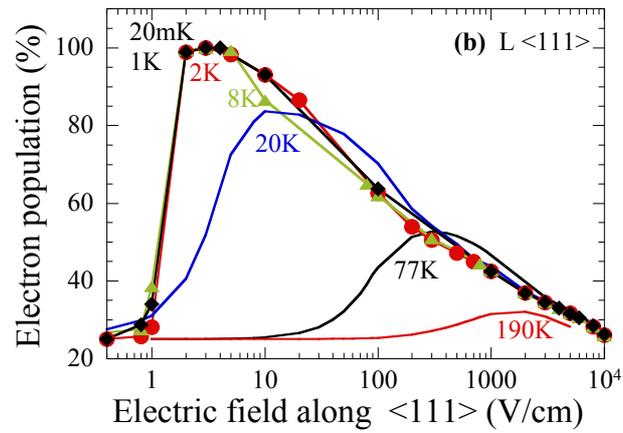

Figure 8

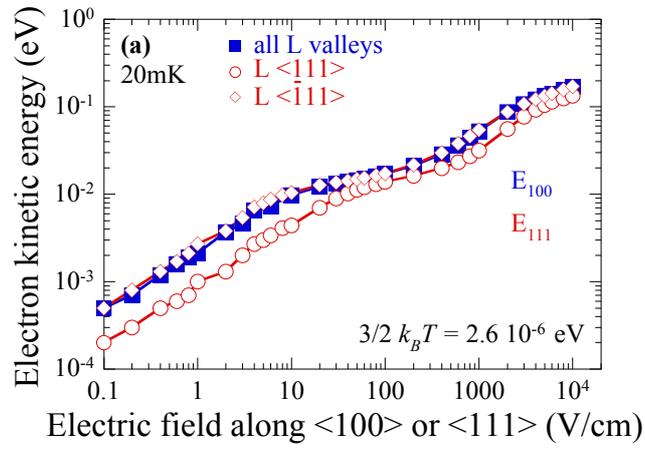

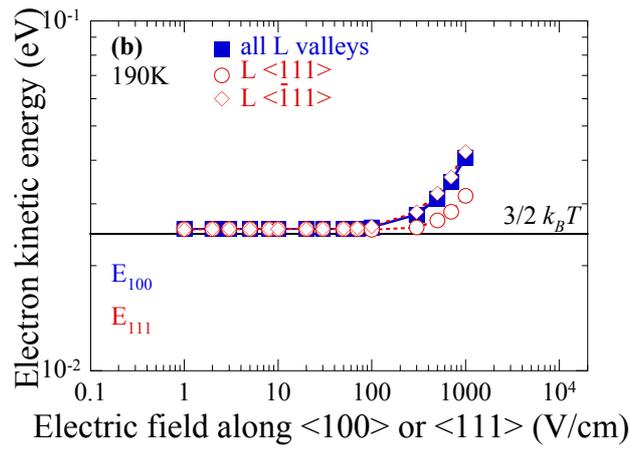

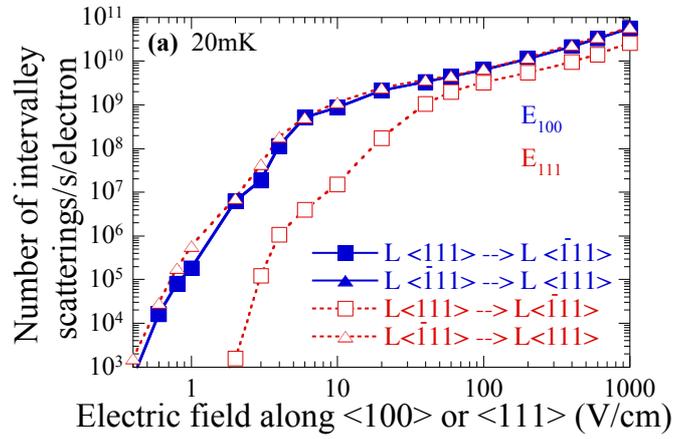

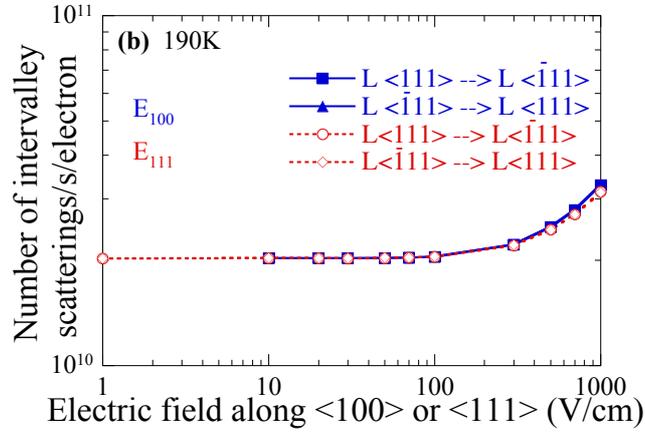

Figure 10

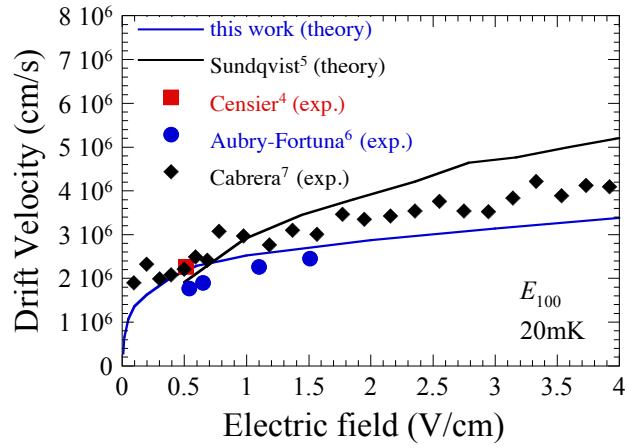

Figure 11

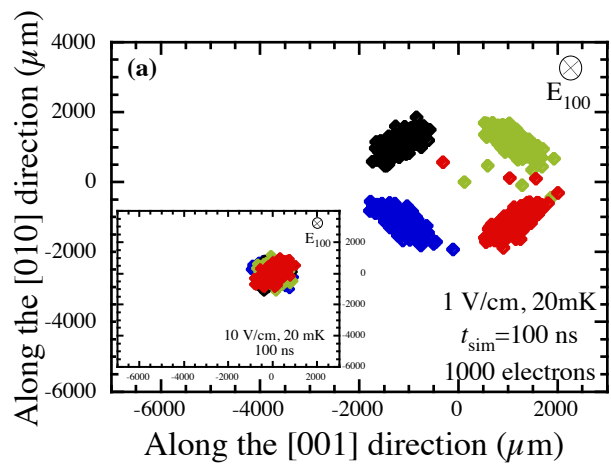

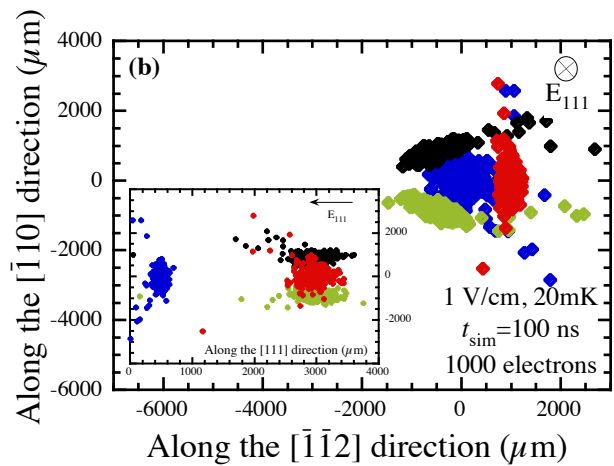